# Nonlinear dynamics of polar regions in paraelectric phase of $(Ba_{1-x},Sr_x)TiO_3$ ceramics

Sina Hashemizadeh[1] and Dragan Damjanovic

*Group for Ferroelectrics and Functional Oxides,*

*Swiss Federal Institute of Technology in Lausanne – EPFL,*

*1015 Lausanne, Switzerland*

The dynamic dielectric nonlinearity of barium strontium titanate $(Ba_{1-x},Sr_x)TiO_3$ ceramics is investigated in their paraelectric phase. With the goal to contribute to the identification of the mechanisms that govern the dielectric nonlinearity in this family, we analyze the amplitude and the phase angles of the first and the third harmonics of polarization. Our study shows that an interpretation of the field-dependent polarization in paraelectric $(Ba_{1-x},Sr_x)TiO_3$ ceramics in terms of the Rayleigh-type dynamics is inadequate for our samples and that their nonlinear response rather resembles that observed in canonical relaxor $Pb(Mg_{1/3}Nb_{2/3})O_3$.

---

[1] E-mail: sina.hashemizadeh@epfl.ch

Short-range polar regions or polar entities in ferroelectric and related materials have been of a great interest for their contribution to the piezoelectric, dielectric and mechanical properties. The generic term "polar entity" will be used here to describe nanometric regions with polar order and may refer to: (i) Känzig regions in $BaTiO_3$;[1,2] (ii) polar nano-regions associated with mixed cations in complex solid solutions such as $(La,Pb)(Zr,Ti)O_3$ and $Pb(Mn_{1/3}Nb_{2/3})O_3$, which appear in the nonpolar phase at Burns temperature;[3] (iii) polar nano-regions embedded in long-range polar domains in e.g. $Pb(Mn_{1/3}Nb_{2/3})O_3$-$PbTiO_3$[4]; (iv) polar boundaries in complex tweed structures in ferroic materials[5,6]; (v) precursors of the ferroelectric phase just above the Curie temperature, $T_C$;[7,8] (vi) defects-induced polar regions appearing above $T_C$,[9] and (vii) residual domains[10], micropolar[11] regions or microregions[12] above $T_C$ in ferroelectric-based compounds. The physics and chemistry of the local polar entities is not fully understood although their contribution to the properties may be significant or even dominant. Examples include the large and frequency dispersive dielectric permittivity in relaxors[13-15], the large piezoelectric effect above ~150 K in relaxor-ferroelectrics[4], the flexoelectric coupling in relaxor-ferroelectrics[10,12,16,17] and the macroscopic polarization in nominally nonpolar phases observed in ferroelectric materials[18].

In this paper we address effects of short-range polar entities on the nonlinear polarisation response in paraelectric phase of $(Ba_{1-x},Sr_x)TiO_3$ (BST) solid solution. BST is technologically interesting for the high tunability of its dielectric permittivity at microwave frequencies. Polar entities, whose presence has been indicated in the paraelectric phase of BST thin films[19] and bulk materials,[20] are of concern because their displacement contributes to the dielectric loss. A broader motivation for the present work is twofold. First, as-prepared paraelectric BST compositions unexpectedly exhibit a small but measurable macroscopic polarization[18,21] possibly arising from alignement of polar entities. Second, the largest flexoelectric coefficients have been reported for this solid solution at temperatures just above Curie temperature, $T_C$[22,23]. The experimental values of the flexoelectric coefficients in this solid solution cannot be recoincilied easily with theoretical predictions.[24-27] Garten *et al.*[10,12,17,28] proposed that in their BST thin films and ceramics the flexoelectric polarization in paraelectric phase is enhanced by polarization from residual ferroelectric domains (or microdomains or micropolar regions), leading to exceptionally large apparent flexoelectric coefficients. Similar result was reported by Narvaez and Catalan for the flexoelectric response in the paraelectric phase of a single crystal relaxor-ferroelectric.[16] Moreover, Garten *et al.* were able to permanently polarize their BST ceramics in the paraelectric phase by simple bending[12]. The macroscopic polarization was explained by a stress-gradient induced alignment of polar entities. The increase of the permittvity in BST films and ceramics with the increasing amplitude of alternating (ac) electric field was interpreted[10,12,17] in terms of the Rayleigh-type dynamics[29,30] of residual domain walls (or another type of interfaces). Interestingly, the macroscopic polarization reported in the paraelectric phase of BST in Ref.[18], has been observed in samples that were never cooled to the ferroelectric phase. In that case, polar entities cannot be linked to a "residual ferroelectricity" but rather considered as "precursors" of the ferroelectric phase[5-8] or are similar to the polar regions discussed in Ref. [3].

To reveal mechanisms that lead to macroscpic polarization and large flexoelectric effect in BST it is therefore important to understand the exact nature and origin of the local polar regions in these materials. Note that origins of the macroscopic polarization and large flexoelectric coefficients do not have to be the same. Such studies should involve in-situ atomic-scale studies to directly establish presence of polar entities[28] and elucidate how they respond to external mechanical and electrical fields. We are presently undertaking such investigations. Useful information on the nature of polar entities can be obtained by studying their nonlinear response to dynamic electric field.[31-35] Whereas most studies and models of the dielectric nonlinearity focus on amplitude of the nonlinear response, we also looked closely at the field dependence of the phase angle of the third harmonic. That information can directly reveal wheather nonlinear contributions exhibit hysteretic character. The phase angle analysis of the higher harmonics is particularly interesting[30,36] to test for the Rayleigh-like dynamics, which was proposed in Refs. [10,12,17] for BST above $T_C$ (or $T_{max}$ in samples that exhibit relaxor behavior). We find that nonlinear polarization in BST ceramics investigated in our study cannot be fully described above $T_C$ by the Rayleigh relations. Those relations hold reasonably well below $T_C$ in soft ferroelectrics where domain walls move in a random energy landscape and where dynamic nonlinearity is essentially hysteretic. In contrast, the nonlinear behavior of BST samples investigated in this study is very similar to that reported in ergodic phase of canonical relaxor $Pb(Mn_{1/3}Nb_{2/3})O_3$ (PMN). [32,34-36]

We report in this letter results for $Ba_{0.6}Sr_{0.4}TiO_3$ (BST60/40), and two end-members of the BST solid solution, $BaTiO_3$ (BTO) and $SrTiO_3$ (STO). The $T_C$ of BST60/40 is 273 K during cooling and at room temperature the material is paraelectric (see Supplementary Information (SI), S1, Fig. S1). Our BTO exhibits $T_C$ of 393 K on cooling and was examined both in ferroelectric and paraelectric state. Its polarization response at room temperature, in the ferroelectric phase with tetragonal structure, is dominated by domain wall motion[37,38]. STO is an incipient ferroelectric with cubic structure at room temperature and its dielectric properties are only weakly dependent on applied electric field at room temperature[39]. We use STO in this paper as a reference quasi-linear dielectric material. For a comparison with BST60/40, we also report on the dielectric nonlinearity in PMN ceramics.

Samples were prepared by mixing BTO and STO powders in desired stoichiometry. Powders were mixed with 4% water-based solution of polyvinyl alcohol (PVA), with binder to powder ratio of 1:25. Powders were pressed in a steel/WC die and the disk-shaped samples were sintered in air at 1723 K for 4 h with a heating rate of 5 K/min and then cooled down by the natural cooling of the furnace. The sintered samples had diameter of about 5.5 mm, thickness of about 0.5 mm and relative density was about 98%. The samples were polished and gold electrodes were sputtered covering completely the major faces. More details on precursors, microstructure and preparation methods can be found in Ref. [18]. Preparation of $Pb(Mg_{1/3}Nb_{2/3})O_3$ ceramics has been described in Ref.[40]. Dielectric nonlinearity was studied by measuring capacitive current of the sample with a sample (capacitance C) placed in series with a standard resistor, R. The driving voltage $V_D(t)=V_{D0}\sin(\omega t)$ with varying amplitude $V_{D0}$ was applied on the circuit. The frequency of the driving field was 1 kHz. To insure that the voltage drop across the sample $\approx V_D(t)$, the value of R was chosen to fulfill condition $R<<1/\omega C$. The driving voltage $V_D$ was generated by a lock-in amplifier and amplified by a wide band amplifier. The capacitive current was determined by measuring voltage on the resistor, $V_R(t)=V_{R0}\sin(n\omega t+\phi)$, where n is the number of the harmonic and $\phi$ is the phase angle. The voltage $V_R$ was measured by the same lock-in amplifier for the first and the third harmonics. The complex permittivity for the nth harmonic was calculated from the complex capacitance whose modulus is given by $C_n=V_{R0}/(V_{D0}Rn\omega)$. Detailed description of the measuring technique can be found in Refs. [36,41,42]. We point out that the behavior described in this and our earlier papers[18] has been observed in dozens of samples prepared from different precursors (e.g. sol-gel, carbonates/oxides, titanates), by different sintering techniques (e.g. spark plasma sintering, conventional sintering) and different laboratories. Some examples are shown in SI S2 Fig. S2. We cannot rule out, however, that samples prepared by another route exhibit a qualitatively different behavior.

In materials with disordered pinning centers for domain walls, the energy potential seen by domain walls is random and the pinning-depinning process is hysteretic and nonlinear.[43-45] A typical example is soft (donor-doped) Pb(Zr,Ti)O$_3$ (PZT).[46] The polarization response may be then well approximated by the following Rayleigh relation:[30,44]

$$P(E) = (\varepsilon_{init} + \alpha E_0)E \pm \frac{\alpha}{2}(E_0^2 - E^2) + \cdots \quad (1)$$

where P is polarization, $E = E_0 \sin(\omega t)$ is applied alternating electric field, $\varepsilon_{init}$ is dielectric permittivity at zero field and α is the Rayleigh coefficient, which describes nonlinearity and hysteresis: sign "+" stands for decreasing and "−" for increasing part of the alternating field. In the case of ferroelectric domain walls, the relation is valid under global subswitching conditions (roughly, for $E_0$ < global coercive field, $E_C$). The relationship (1) can be expanded into Fourier series yielding:

$$P(E) = (\varepsilon_{init} + \alpha E_0)E_0 \sin(\omega t) - \frac{4\alpha E_0^2}{3\pi}\cos(\omega t) - \frac{4\alpha E_0^2}{\pi}\left[\frac{1}{15}\cos(3\omega t) - \frac{1}{105}\cos(5\omega t) + \cdots\right] \quad (2)$$

The key feature of equation (1) is that all higher harmonics are out-of-phase with the driving field, meaning that they contribute to both the hysteresis and nonlinearity. This hysteresis-nonlinearity relationship will be analyzed next. It is important to understand that Eq. (1) may contain additional terms to better describe a real, non-ideal material,[29] and that the main feature of the equation is the link between the nonlinearity and hysteresis (as seen in Eq. (2)). Additional terms in (1) reflect degree of randomness of the energy profile.[44-46] As long as this link between the hysteresis and nonlinearity is present the system is referred to as Rayleigh-like.[46] The absence of even harmonics in Eq. (2) is a consequence of the assumption that the system is (ideally) symmetric with respect to the driving field direction, which is only an approximation for a real sample [32,41,42,47] (See also SI S3). In principle, Eq. (1) may be valid for other type of interfaces, not only for domain walls. It should be also understood that domain walls in ferroelectrics may exhibit (and often do) a different dynamics from that described by Eq. (1). A good example is acceptor-doped, hard PZT which in its well aged state cannot be described by Rayleigh relations but can in de-aged state.[41,42,48]

Fig. 1 shows the results of harmonic analysis of a BTO sample at room temperature. As may be expected for a relatively soft ferroelectric material, the field dependence of the real part of the permittivity defined as $\varepsilon_{ac}(E_0)=P(E_0)/E_0$, suggests the Rayleigh-like behavior (Eq. 1) at fields above ≈ 0.35 kV/cm, Fig. 1(a). The change of the slope below this threshold field signifies that the energy profile is not perfectly random.[46] Because of a low coercive field for this BTO sample ($E_C$≈1.5 kV/cm), the field dependence of the permittivity is nontrivial and the investigation here is focused on the nonlinear hysteretic behavior of the material, i.e. on the phase of the third harmonic. A presence of a Rayleigh-like mechanism can be verified by the value of the phase angle of different harmonics. Fig. 1(b) shows that at larger fields the phase angle of the 1$^{st}$ harmonic ($\delta_1$) is close to zero while the phase angle of the 3$^{rd}$ harmonic ($\delta_3$) is close to -90°, as predicted by Eq. (1). Note that the quadrature (out-of-phase) component of the first harmonic is non-zero, but its amplitude is relatively small with respect to the in-phase component, leading to a nonzero but small $\delta_1$ (see Fig. S1 in SI for details). The value of the $\delta_3$, roughly around -90°, indicates that the nonlinear motion of domain walls is at the same time strongly hysteretic.

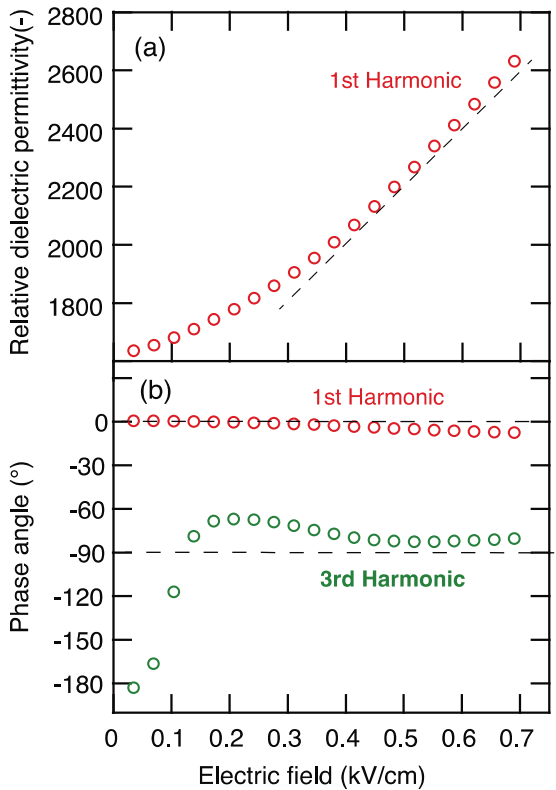

Figure1. (color online). (a) The relative dielectric permittivity (from the 1$^{st}$ harmonic) and (b) phase angle for the 1$^{st}$ and the 3$^{rd}$ harmonics of the polarization for a BTO sample. The dashed line in (a) is a guide to eye to indicate transition into linear, Rayleigh-like regime. The horizontal dashed lines in (b) indicate values expected from the ideal Rayleigh-like behavior, as predicted by Eq. (1). The coercive field for this sample is about 1.5 kV/cm.

The rapid increase of $\delta_3$ from -180° to ~-90° is in agreement with the threshold field observed in Fig. 1(a) and may suggest non-uniformity in distribution of potential energy barriers (i.e., the barriers' heights may not extend uniformly all the way to zero) [30]. The value of the phase angle of ~ -180° at the weak fields indicates nearly anhysteretic behavior at weak fields, and is consistent with the displacement of domain walls in potential wells that are too deep to be overcome by the applied electric field. Once the field is strong enough, the domains are able to depin from defects and move in hysteretic, nonlinear, irreversible fashion, indicated by the phase angle of $\delta_3$ approaching -90°. At fields higher than those shown in Fig. 1, the large-scale switching events lead to saturation of the lock-in amplifier.

Fig. 2 shows harmonic analysis of an STO ceramic. The relative dielectric permittivity determined from the 1st harmonic and the phase angles of the 1st (≈0°) and the 3rd (-180°) harmonics are nearly independent of electric field. Therefore, the first and the small third harmonic both contribute to the polarization in essentially anhysteretic fashion (a small hysteresis is present because the angles are not perfectly 0° or -180°). If STO contains any polar entities,[49] their contribution to polarization is small and their dynamics cannot be described by Eq. (1); it is neither (strongly) nonlinear nor hysteretic at these fields and at room temperature.

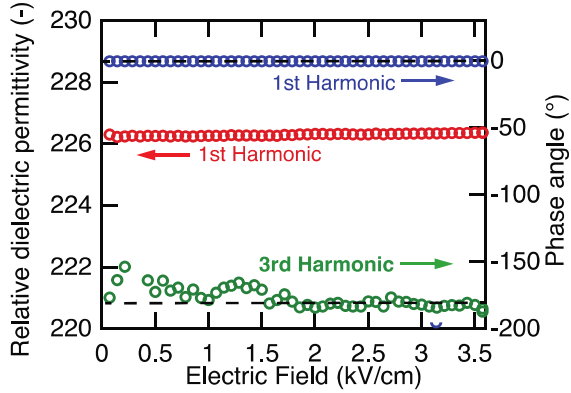

Figure2. (color online). Real part of the relative dielectric permittivity for the first harmonic and the phase angles of the 1st and the 3rd harmonics for STO. Dashed lines represent the expected values for phase angles of the 1st and the 3rd harmonics for an ideal linear, anhysteretic material.

We next look at BST60/40 and discuss the main result of this work. The breaking of the cubic symmetry in the paraelectric phase of this material is confirmed by pyroelectric measurements (representative data can be found in Ref. [18]). It has been suggested in Ref.[18], that polar entities could be biased by a strain gradient resulting in the breaking of the centric macroscopic symmetry in the paraelectric phase. Similar idea was also proposed by Garten *et al*. for BST thin films and ceramics.[10,12,17,50] They demonstrated that a strain gradient may polarize BST samples and suggested that it happens by reorientation of residual ferroelectric domains (or another kind of polar entities) whose dynamics can be described by Eq. 1. To get more information on the nature of these polar objects, we measured and analyzed nonlinear polarization response in our BST60/40 samples considering the following two verifiable possibilities: (i) the dynamic of polar objects is similar to that of domain walls in soft ferroelectrics, resulting in the Rayleigh-like behavior of polarization.[10,12,50] If that is the case, the response should be qualitatively similar to that in BTO sample shown in Fig. 1 and should exhibit nonlinearity and hysteresis in all higher odd harmonics (ideally non-zero quadrature and zero in-phase components of polarization); and (ii) the dynamics of local polar entities leads to a nonlinear but nearly anhysteretic response (ideally zero quadrature and non-zero in-phase components of higher harmonics of polarization), as suggested in Ref.[36] for response of polar-entities in the ergodic phase of $Pb(Mg_{1/3}Nb_{2/3})O_3$.

Harmonic analysis of the polarization as a function of electric field amplitude was therefore carried out for several BST60/40 ceramics and, for comparison, for a PMN ceramic. The permittivity data for the first harmonic and $\delta_3$ are shown for a BST60/40 and a PMN sample in Fig. 3. In SI S2 and Fig. S2, we discuss and show a similar set of data for different BST60/40 samples.

The first interesting observation is that the nonlinear behavior of different BST60/40 and PMN samples is qualitatively similar (compare PMN data in Fig. 3 and data for a BST60/40 sample with similar microstructure in Fig. S2): at low fields, for the majority of the samples, the permittivity increases with increasing field, followed by a negative nonlinearity with increasing field for all samples. Exactly the same ac field dependence of the permittivity has been reported for thin films of PMN and BST.[32,33] It is significant that thin films of BST in Ref.[50] exhibit relaxor behavior, whereas our ceramics (SI S1, Fig. S1) behave as "normal" ferroelectrics. Yet, their behaviors above $T_{max}$ (films) and $T_C$ (ceramics) are similar.

The decrease in the permittivity at high fields could be related to two mechanisms: one is "tunability" of the permittivity[39] and the other is reorientation of polar entities followed by saturation of this response at high fields.[32,34,51,52] Tunability usually refers to the dependence of the intrinsic, lattice polarization response on static field and ensuing decrease of the permittivity with an increasing field strength. A simple phenomenological model of a nonlinear, centrosymmetric dielectric shows that a qualitatively similar behavior of the dielectric nonlinearity may be derived for alternating and static fields.[34,51,52] On the basis of the present data alone, therefore, we cannot conclude what is the origin of the negative nonlinearity at higher fields in these samples.[34,39]. While the two mechanisms can act concurrently, the previous studies favor interpretation in terms of a dominant contribution from orientable polar entities at high fields.[32-34,39] Our data are consistent with this interpretation.

The situation is in general clearer at weak fields, where contribution of polar entities (for example, from displacement of domain walls, or "breathing" of polar regions) dominates polarization response[34,39]. The permittivity then usually, but not exclusively[34], increases with an increasing field amplitude. The initial increase of the permittivity in our samples with increasing field, thus, suggests presence of some kind of electrically active polar entities above $T_C$.

We now look at the phase angle of BST60/40 samples in the two main regimes that can be identified in Fig. 3 and Fig.S2 in SI. The $\delta_3$ globally approaches either ≈ 0° or ≈-180°, indicating that the amplitude of the quadrature component of the nonlinear polarization for this harmonic is close to zero. This is reminiscent of the behavior previously suggested for PMN[36] and confirmed experimentally in this work (see Fig. 3). The physical meaning of values of $\delta_3$= 0° and -180° is the following.[41] When $\delta_3$ ≈-180°, the third harmonic increases the amplitude of the total polarization. When $\delta_3$ ≈ 0°, the third harmonic decreases the response amplitude. Both effects are ideally anhysteretic and agree well with the trend in the permittivity which first increases and decreases when $\delta_3$ switches towards zero. The evolution of the $\delta_3$ and permittivity with the field amplitude are due to a transition between two mechanisms, none of which corresponds to prediction of Eq. (1), which requires that $\delta_3$ be close to -90°. The transition region from -180° to 0° and absence of a Rayleigh-like regime at any filed range is also seen in Fig. S4 (SI S4) which shows $\delta_3$ evolution with field in several BST compositions.

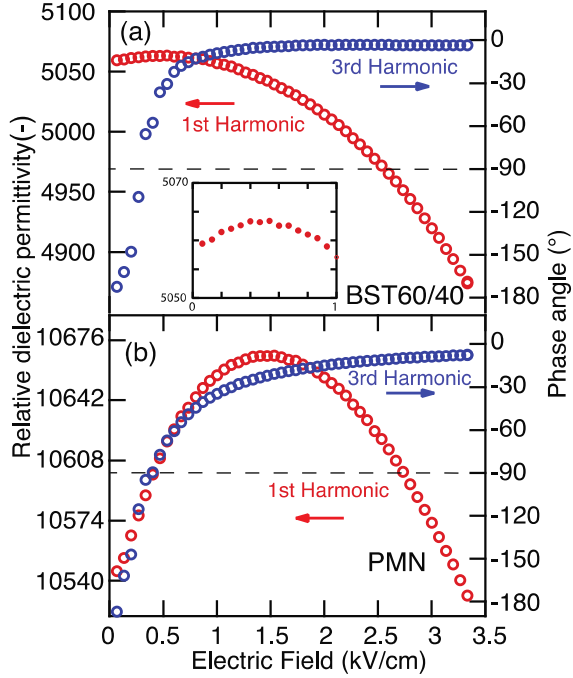

Figure 3. (color online). The real part of the relative dielectric permittivity measured at the first harmonic and the phase angle of the third harmonic for: (a) a BST60/40 ceramic sample with no history of the ferroelectric phase transition and (b) a PMN ceramic. Dashed lines represent expected values for the phase angle of the 3$^{rd}$ harmonic for a material with the Rayleigh-like behavior. Inset in (a) shows the maximum in permittivity at low fields. See Fig. S4 in SI for response of other BST60/40 samples. All measurements were made at room temperature.

We also examined response of BTO samples above $T_C$. As BST, BTO possesses macroscopic polarization[18] above $T_C$ which might indicate presence of polar entities.[53] Fig. 4 shows that twenty degrees above $T_C$ a BTO sample exhibits nonlinear permittivity which increases with increasing field, and the $\delta_3$ evolves with the field in a similar fashion as in BST60/40 and PMN: it rapidly changes from ~ -180° toward ~ 0° at weak fields and then shows tendency to stabilize toward 0° at large fields. We speculate that this trend in $\delta_3$ might anticipate a maximum and then a decrease in the permittivity at higher fields than used here. If so, this could be an indication that the nonlinear response described in this letter is not exclusively due to chemical inhomogeneities at cation sites. Finally, over the examined field range the BTO sample does not exhibit dielectric nonlinearity that can be described by Eq. (1).

In summary, we have analyzed dependence of the nonlinear dielectric properties on the amplitude of the ac electric field in the paraelectric phase of ferroelectric BST60/40 and BTO ceramics. The nonlinear response, with a low hysteresis in the third harmonic over most of the examined field range, is similar to that observed in relaxor PMN ceramics and is qualitatively different from the one that can be described by a Rayleigh-like mechanism. The results presented here do not imply that the nonlinear dynamic behavior of BST60/40 and BTO above $T_C$ and of PMN above $T_{max}$ cannot be described by a domain-wall like dynamics. What these results do show is that the dynamics of polar entities in these materials cannot be well described in its totality by the Rayleigh-like relations that are otherwise valid for description of domain-wall contributions in soft ferroelectrics, such as donor-doped PZT. The nonlinear behavior reported here for the permittivity and $\delta_3$ can thus serve as a test for validity of models of dynamics of polar entities in the paraelectric phase of ferroelectrics.

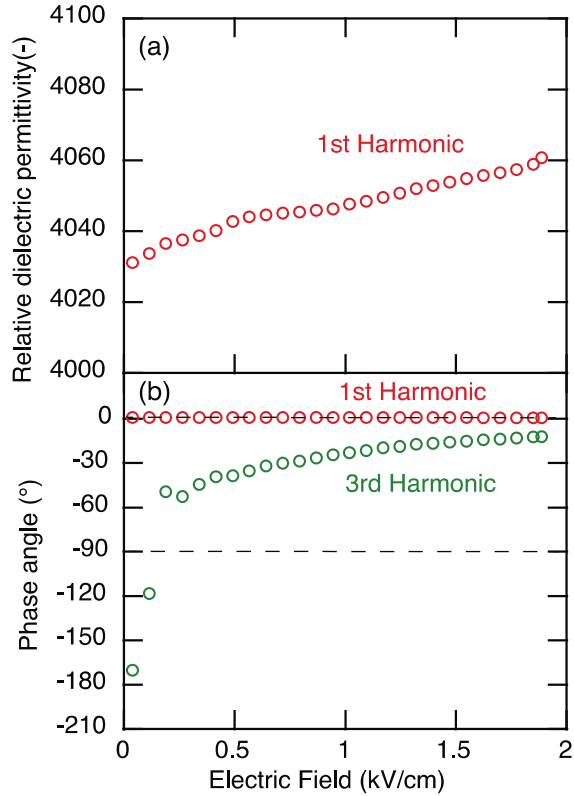

Figure4. (color online) (a) The real part of the relative dielectric permittivity measured at the first harmonic and (b) the phase angle of the first and the third harmonic for a BTO sample. All measurements were made above $T_C$, at 413 K. Dashed lines represent the expected values for the phase angles of the $1^{st}$ and $3^{rd}$ harmonics of a material with the Rayleigh-like behavior.

**Supplementary Material**

Supplementary material includes additional data and discussion on: 1) dielectric permittivity, loss and P-E loops in BST60/40, 2) dielectric nonlinearity in BST60/40 samples with different history and from different sources, 3) influence of second harmonic on the data, and 4) dielectric nonlinearity for BST samples with different barium content.

**Acknowledgements:** This work was supported by the Swiss National Science Foundation (No. 200021-159603). Samples of PMN were kindly supplied by Hana Uršič and some BST60/40 samples by T. Hoshina.

# Supplementary Information for
## "Nonlinear dynamics of polar regions in paraelectric phase of $(Ba_{1-x},Sr_x)TiO_3$"
by S. Hashemizadeh and D. Damjanovic

**Supplementary information S1**:

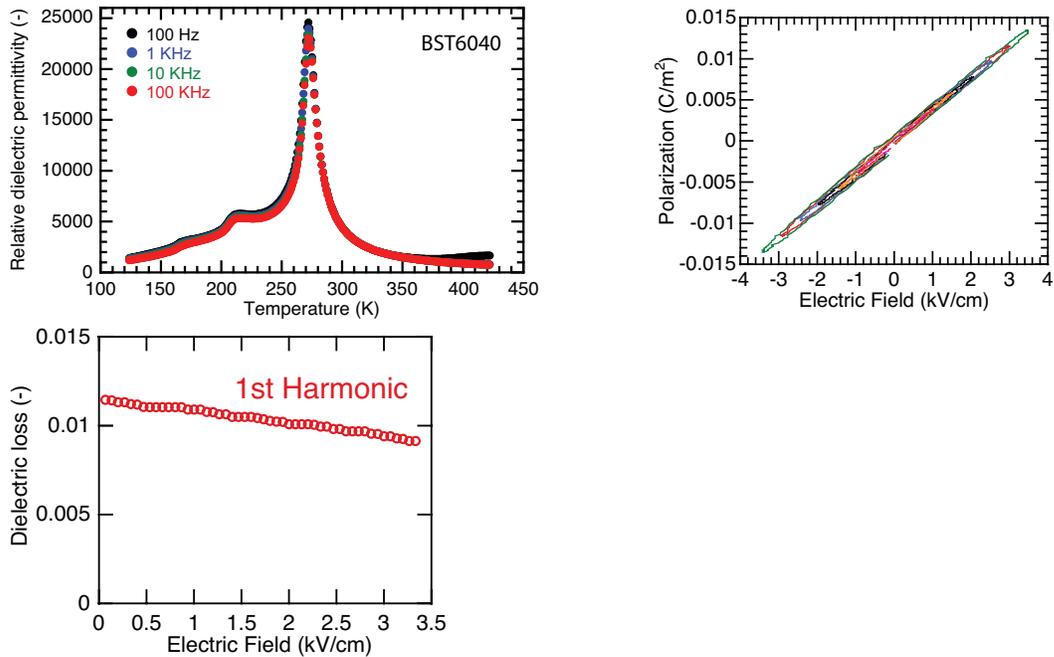

Fig. S1: (upper left) typical temperature dependence of the dielectric permittivity of our BST60/40 samples; (upper right) typical P-E loops for a BST 60/40 samples at room temperature (50 Hz). The slim loops indicate low leakage losses even at highest ac fields; (bottom) dielectric loss of a BST60/40 sample (1 kHz) at different driving fields, indicating a low loss which is decreasing with increasing ac field amplitude. This is the same sample whose other nonlinear data are shown in Fig. 3a.

The sharp peak in permittivity vs temperature function indicates good chemical homogeneity of the samples. A detailed chemical analysis using TEM shows that inhomogeneity in Ba/Sr distribution is limited to nano-sized regions.

**Supplementary information S2:**

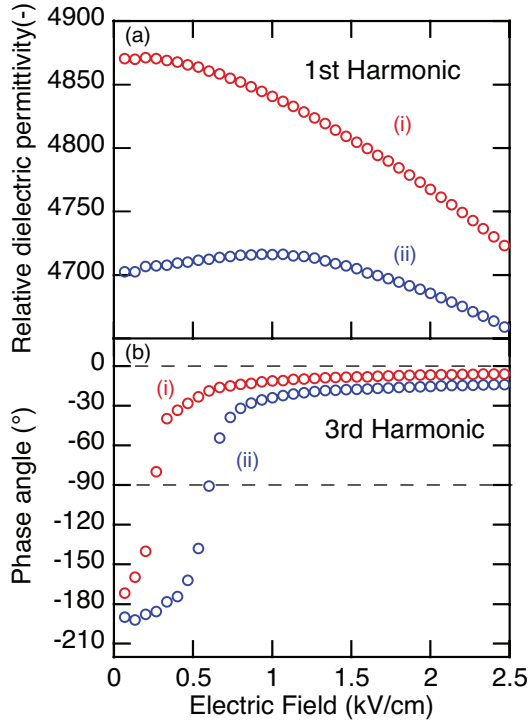
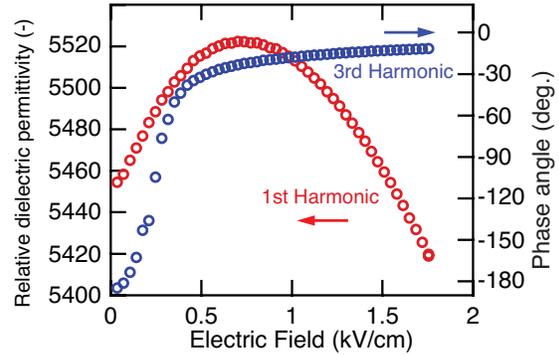

Fig. S2. Illustration of variations in properties of BST60/40 samples with different histories and preparation conditions. All samples show qualitatively similar nonlinear behavior. Compare with data shown in Fig. 3 for BST60/40 and PMN.

Sample marked (i) in Fig. S2 has been prepared in the same way as sample (ii), but sample (ii) has been cooled below $T_C$ to 263 K and then heated up again to room temperature and measured. Sample on the right side in Fig. S2 has been prepared at Tokyo Institute of Technology, by T. Hoshina. It has not been cooled through $T_C$. The main difference between our samples and those prepared by T. Hoshina, is that Hoshina's sample has been sintered at 1130 °C-1220 °C after a rapid heating to 1350 °C. The grain size of Hoshina's sample is on the order of 2-4 $\mu$m, while our samples exhibit grain size >20 $\mu$m.[1] Thus, while the microstructure and history affects the nonlinear behavior, it remains qualitatively similar for all samples. A close inspection of sample (i) shows that it probably exhibits a maximum in the permittivity at weak fields, which is, however, lost in the noise in the data.

It is particularly instructive to compare similarities between the data for the sample prepared by T. Hoshina, Fig. S2 (right), and PMN sample shown in Fig. 3.

**Supplementary information S3:**

A real sample is usually not completely free of the second harmonic in its dielectric response to electric field even if the material is centrosymmetric. This has been discussed in Refs. 2,3 and 4. (see for example Figs. 4.8 and 4.17 in Ref.[5]). The delicate point with the dielectric measurements under applied electric field is that the very first application of the field may disturb the symmetry of the sample. The effect of the $2^{nd}$ harmonic on the nonlinear dielectric behavior may be easily detected either by the Fourier analysis of the polarization signal or, if the $2^{nd}$ harmonic is particularly strong, by observing asymmetry in the P-E loop. Any mechanism that leads to the appearance of the second harmonic will naturally affect the shape of the energy landscape for contributing polar entities. It turns out that in poled ferroelectric films and ceramics which are strongly asymmetric and which exhibit Rayleigh-like behavior, the amplitude of the second harmonic may be comparable to that of the third harmonic[5] but it still does not affect qualitatively the phase angle of the third harmonic, which remains roughly around -90°. (This is so because the asymmetry ideally changes only the quadrature component of the third harmonic but not the in-phase component, which remains zero). In our experience, a relatively small second harmonic does not disturb observation of the underlying Rayleigh law if the analysis is done using Fourier analysis so that the phase angle of each harmonic can be separated.

We do not think that for the present qualitative analysis it is important to consider effects of the second harmonic as long as the experimental behavior confirms (or disproves) qualitatively one of the models and one knows that the deviation due to $2^{nd}$ harmonic cannot be so large as to bring about confusion between different interpretations. See Refs. [4-7] where this has been discussed in some detail.

In our BST 60/40 samples, which exhibit macroscopic polarization in paraelectric phase we observe second harmonic which is about 3 orders of magnitude lower than the amplitude of the first harmonic and is comparable to the amplitude of the third harmonic. Based on our experience with PZT, we do not have evidence that the presence of the second harmonic affects qualitatively analysis presented in this paper.

**Supplementary information S4:**

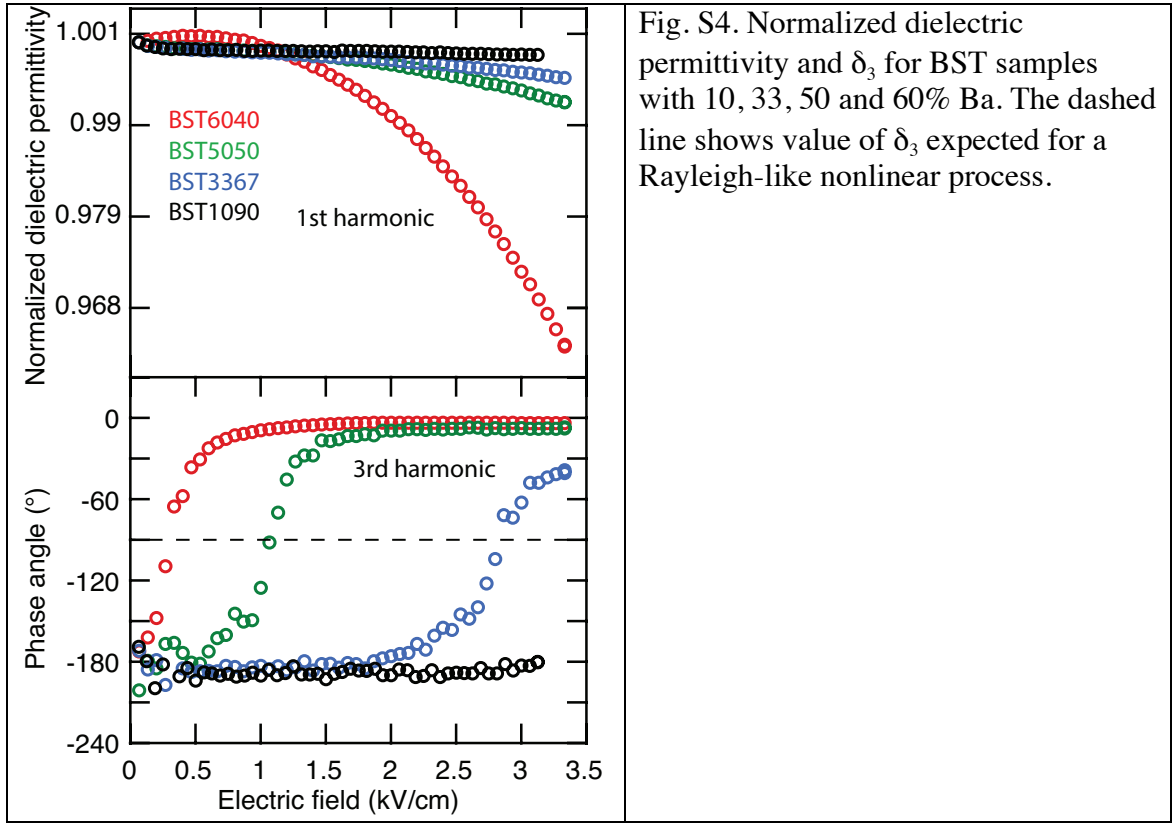

Fig. S4. Normalized dielectric permittivity and $\delta_3$ for BST samples with 10, 33, 50 and 60% Ba. The dashed line shows value of $\delta_3$ expected for a Rayleigh-like nonlinear process.

**References for SI:**